\documentstyle[12pt]{article}

\newbox\SlashedBox  
\def\fs#1{\setbox\SlashedBox=\hbox{#1} 
\hbox to 0pt{\hbox to 1\wd\SlashedBox{\hfil/\hfil}\hss}{#1}} 
\def\hboxtosizeof#1#2{\setbox\SlashedBox=\hbox{#1} 
\hbox to 1\wd\SlashedBox{#2}} 

\def\ms#1{\setbox\SlashedBox=\hbox{$#1$}
\hbox to 0pt{\hbox to 1\wd\SlashedBox{\hfil/\hfil}\hss}#1}

\newcommand{\tr}{{\rm tr}}
\newcommand{\ie}{{\em i.e.~}}
\newcommand{\eg}{{\em e.g.~}}
\newcommand{\be}{\begin{equation}}
\newcommand{\ee}{\end{equation}}
\newcommand{\ba}{\begin{eqnarray}}
\newcommand{\ea}{\end{eqnarray}}
\begin{document} 

\thispagestyle{empty}

\begin{flushright}
ROM2F/99/9
\end{flushright}

\vspace{1.5cm}

\begin{center}

{\LARGE {\bf On the logarithmic behaviour \\ 
in ${\cal N}$=4 SYM theory\rule{0pt}{25pt} }} \\
\vspace{1cm} {\large Massimo Bianchi, Stefano Kovacs, Giancarlo Rossi and
Yassen  S. 
Stanev$^{\dagger}$} \\ 
\vspace{0.6cm} 
{\large {\it Dipartimento di Fisica, \ Universit{\`a} di Roma \  
``Tor Vergata''}} \\  {\large {\it I.N.F.N.\ -\ Sezione di Roma \ 
``Tor Vergata''}} \\ {\large {\it Via della Ricerca  Scientifica, 1}} 
\\ {\large {\it 00173 \ Roma, \ ITALY}} \\

\end{center}

\vspace{1cm}

\begin{abstract}

We show that the logarithmic behaviour seen in perturbative and 
non perturbative contributions to Green functions of gauge-invariant composite
operators in ${\cal N}$=4 SYM with $SU(N)$ gauge  group can be consistently
interpreted in terms of anomalous  dimensions of unprotected operators in long
multiplets of the  superconformal group $SU(2,2|4)$. In order to illustrate the
point we analyse the short-distance behaviour of a particularly simple
four-point Green function  of the lowest scalar components of the ${\cal 
N}$=4
supercurrent multiplet. Assuming the validity of the Operator Product
Expansion, we are able to reproduce the known value of the one-loop anomalous
dimension  of the single-trace operators in the Konishi supermultiplet. We
also show that it does not receive any non-perturbative contribution from the  
one-instanton sector. We briefly comment on double- and multi-trace 
operators and on the bearing of our results on the AdS/SCFT correspondence.   
\end{abstract}

\vspace{4cm}
\noindent
\rule{6.5cm}{0.4pt} 

{\footnotesize ${}^{\dagger}$~On leave of absence from Institute for 
Nuclear Research and Nuclear Energy, Bulgarian Academy of Sciences, 
BG-1784, Sofia, Bulgaria}
\newpage 

\setcounter{page}{1}

\section{Introduction and summary of the results}

As is well known, conformal invariance puts tight 
constraints on the correlation functions of primary operators~\cite{fgg}.
In particular, two- and three-point Green functions are fixed up to 
multiplicative constants.
Four-point functions in four-dimensional 
Conformal Field Theories (CFT's), however,  are determined by 
conformal invariance
only up to {\it a priori}  unknown functions of two independent conformal
invariant  cross-ratios. As a consequence the comparison of four-point
amplitudes computed in  type IIB supergravity/superstring theory in
$AdS_{5}\times S^{5}$  with four-point correlation functions in 
${\cal N}$=4 super Yang--Mills theory (SYM) represents a  truly 
`dynamical' and highly non-trivial check of the correspondence 
proposed by Maldacena~\cite{jm}. 

A dynamical check of the Maldacena conjecture was indirectly carried 
out for some higher-derivative terms in the AdS effective action in~\cite{bgkr}, 
where even in the case of an $SU(2)$ gauge group the non-perturbative 
(one-instanton) correction to a sixteen-fermion Green function was shown 
to precisely match 
the result of a similar non-perturbative (one-D-instanton) correction 
in the AdS supergravity theory. A supersymmetry related four-point 
function of scalar composite operators belonging to the ${\cal N}$=4 
supercurrent multiplet was also explicitly computed. This matching was
recently proven to hold for any number of colours, $N$, and for any 
instanton number in the large $N$ limit~\cite{dhkmv}.

A common feature of all four-point Green function computations performed so
far, both in the AdS  description~\cite{freed4point1,brodiegut,liu,dfmmr} and
in ${\cal N}$=4 SYM theory~\cite{bk,ehssw}, is the appearance of 
short-distance logarithmic singularities. In this paper we demonstrate that
in the conformal phase of SYM theory this behaviour 
can be explained as a result of the 
presence of operators with non-vanishing anomalous dimensions.

Indeed only short supermultiplets of the global $SU(2,2|4)$ superconformal
symmetry are protected from receiving quantum  corrections to their scale
dimensions, $\Delta$~\cite{af}. 
The lowest scalar components of a short multiplet belong to a  
representation of highest weight $[0,\ell,0]$ of the $SU(4)$ R-symmetry
and have scale dimension
$\Delta=\ell$, for some integer  $\ell\geq 2$~\cite{gunmar}.
A familiar example is the ${\cal N}$=4 supercurrent multiplet with $\ell=2$,
whose AdS  counterpart is the gauged supergravity multiplet.

Long multiplets on the contrary can exist for any value of the scale
dimension of their lowest component~\cite{af}.
An example is the Konishi 
supermultiplet~\cite{koni} that involves single (colour) trace
operators. In the large $N$ limit the Konishi-type 
supermultiplets, that  correspond 
to genuine string excitations~\cite{af}, are expected to receive 
(large) corrections to their anomalous dimensions~\cite{gkp,wittads}.

There exist also multi-trace operators that may belong both 
to short and to long  multiplets~\cite{af}. The former are protected 
while, as we will show in this paper, 
the corrections to the anomalous dimensions of 
the latter are vanishingly small in the large $N$ limit.

In order to illustrate our point, in this paper we  concentrate on the 
simplest example of unprotected operator, namely the lowest scalar
component of the Konishi multiplet,
which has 
naive scale dimension $\Delta^{^{(0)}}=2$.
We compute both the one-loop and the one-instanton contributions
to a particular four-point function in which this operator is exchanged 
in only one of the intermediate channels.
Assuming the validity of the
Operator Product Expansion (OPE),
we reproduce the known value of the one-loop 
anomalous dimension of the  Konishi multiplet as a
non trivial check of our interpretation of the appearence of short-distance
logarithmic singularities. 
Moreover, the analysis of the non-perturbative
contribution demonstrates that both the
anomalous dimension of the Konishi multiplet and its trilinear OPE
coefficients with  operators in the supercurrent multiplet receive only
perturbative corrections.

The outline of the paper is as follows. After a preliminary 
 discussion on
anomalous  dimensions in a generic CFT in section 2, in section 3 we identify
a simple four-point  function that allows us to most directly expose 
and interpret the short-distance 
logarithmic behaviour in ${\cal N}$=4 SYM theory. One-loop and one-instanton
computations are described  in sections 4 and 5, respectively.
Section 6 contains our conclusions and comments on
the logarithmic behaviour found in genuine AdS computations.

\section{Anomalous dimensions in Conformal Field Theory}

Anomalous dimensions usually (\eg in QCD) appear as a consequence of 
the need of regularising and renormalising the theory. At first sight 
their appearance might seem surprising in a theory, such as $SU(N)$ ${\cal
N}=4$ SYM, that is `known' to be finite~\cite{stef}. Computations 
performed so far have indeed shown that four-point functions of composite
operators belonging to the supercurrent multiplet are finite at 
non-coincident points. On the other hand, even 
two-point functions of protected operators, that are expected not 
to be corrected in perturbation theory nor after inclusion of instanton
effects, are not finite in a distributional sense. 
In order to make them finite  one  has to subtract short-distance 
non-integrable 
singularities~\cite{chalmschalm2}. This,
however,  is necessary even in a free theory and
has nothing to do with  the non-trivial dynamics of $SU(N)$ ${\cal N}$=4 SYM. 

To illustrate  in a simple fashion the emergence of logarithmic terms   
and their relation to anomalous dimensions,
let us consider the two-point function of a primary
 operator of scale dimension $\Delta$ 
\begin{equation}
\langle {\cal O}^\dagger_\Delta (x) {\cal O}_\Delta (y) \rangle   = 
{A_\Delta 
\over 
(x-y)^{2\Delta}} \; ,
\label{exact}
\end{equation}
where $A_\Delta$ is an overall normalisation constant possibly 
depending on the subtraction scale $\mu$. Now suppose that 
$\Delta = \Delta^{^{(0)}} + \gamma$, \ie the operator under consideration 
has an anomalous dimension.  In  perturbation theory $\gamma=\gamma (g)$ 
is expected to be small and to admit an expansion  in the
coupling constant $g$. The perturbative expansion of~(\ref{exact}) 
in powers of $\gamma$ yields  
\begin{equation}
\langle {\cal O}^\dagger_\Delta (x) {\cal O}_\Delta (y)\rangle
= { a_\Delta \over (x-y)^{2\Delta^{^{(0)}}}} \left( 1 - \gamma \log 
[\mu^2 (x-y)^2] + {1 \over 2} \gamma^2 
( \log [\mu^2 (x-y)^2] )^2 + \ldots \right) \; ,
\label{defdelta}
\end{equation}
where after renormalisation 
we have set $A_\Delta = a_\Delta \mu^{-2\gamma}$.
Thus, although the exact expression~(\ref{exact})
is conformally invariant, 
at each  order in $\gamma$~ (or in $g$) (\ref{defdelta}) contains 
logarithms that seem to even violate scale invariance. 
Let us stress that 
the appearance of logarithmic terms is
an artifact of the 
perturbative expansion, rather than an intrinsic property of the
theory.

Similar considerations apply to generic  
$n$-point Green functions as well. 
Assuming the validity of the OPE, a four-point 
function can be expanded in the $s$-channel in the form
\begin{equation}
\langle {\cal Q}_{A}(x) {\cal Q}_{B}(y) {\cal Q}_{C}(z) {\cal Q}_{D}(w) 
\rangle  = 
\sum_K {C_{AB}{}^K (x-y,\partial_{y})\over (x-y)^{\Delta_A + 
\Delta_B-\Delta_K}}
{C_{CD}{}^K (z-w,\partial_{w})\over (z-w)^{\Delta_C + 
\Delta_D-\Delta_K}} 
\langle {\cal O}_{K}(y) {\cal O}_{K}(w) \rangle \; ,
\label{doubleope}
\end{equation}
where $K$ runs over a (possibly infinite) 
complete set of primary operators. Descendants 
are implicitly taken into account by the presence of derivatives in the 
Wilson coefficients, $C$'s. An expansion like~(\ref{doubleope}) is valid 
in the other two channels  as well. To simplify formulae we assume that 
${\cal Q}_{A},{\cal Q}_{B},{\cal Q}_{C},{\cal Q}_{D}$ are protected 
operators,~\ie they have no 
anomalous dimensions. In general the operators ${\cal O}_{K}$ may 
have anomalous dimensions, 
$\gamma_K$, so that $\Delta_K=\Delta^{^{(0)}}_K + \gamma_K$, where 
$\Delta^{^{(0)}}_K$ is the tree-level scale dimension. Similarly 
$C_{IJ}{}^K = C^{^{(0)}}_{IJ}{}^K + \eta_{IJ}{}^K$, with 
$\eta_{IJ}{}^K$ the perturbative correction to the OPE coefficients.
Indeed, although three-point functions of single-trace chiral primaries 
are known not to be renormalised beyond tree level~\cite{dfs}, 
{\it a priori} nothing can be said concerning corrections to three-point
functions involving also unprotected operators.

Neglecting descendants, one gets for the first-order terms 
of~(\ref{doubleope}) in the small parameters $\gamma$ and $\eta$ 
\begin{eqnarray}
    && \langle {\cal Q}_{A}(x) {\cal Q}_{B}(y) {\cal Q}_{C}(z) 
    {\cal Q}_{D}(w) \rangle_{_{(1)}} = \sum_K
    {\langle {\cal O}_{K}({y}) {\cal O}_{K}({w}) 
    \rangle_{_{(0)}} \over (x-y)^{\Delta_A + 
    \Delta_B-\Delta^{^{(0)}}_K} (z-w)^{\Delta_C + 
    \Delta_D-\Delta^{^{(0)}}_K}} \cdot \nonumber \\
    && \cdot \left[ 
    \rule{0pt}{18pt}\eta_{AB}{}^K 
    C^{^{(0)}}_{CD}{}^K + C^{^{(0)}}_{AB}{}^K \eta_{CD}{}^{K} + 
     {\gamma_K \over 2} C^{^{(0)}}_{AB}{}^K 
    C^{^{(0)}}_{CD}{}^K \log {(x-y)^2 (z-w)^2 \over 
    (y-w)^4 }\right] \; .
    \label{opex}
\end{eqnarray}
From this perturbative formula one can extract the corrections
to both the OPE coefficients and the anomalous dimensions
of the operators ${\cal O}_{K}$. The 
former come from the terms in~(\ref{opex}) that display the same 
singularities as dictated by naive dimensional analysis, the latter 
from the coefficients of the logarithmic terms.

We end this section with a general remark on anomalous dimensions.
Positivity of two point-functions in the distributional sense 
implies lower bounds on the anomalous dimension of the operators. In 
particular, scale dimensions of rank $r$ symmetric tensors have 
to satisfy  $\Delta \ge 2 + r$~\cite{dmppt}. For a scalar field the 
bound is $\Delta \ge 1$, which is obviously saturated by 
a free field. If, however, the scalar field belongs to a 
supermultiplet that also contains some tensor field, there are  
additional constraints coming from the relation imposed by 
supersymmetry between the scale dimensions of scalar and tensor 
fields~\cite{ans}. 

\section{Tree-level Considerations}

In this section we identify a simple 
four-point function of protected operators that involves the exchange 
of unprotected operators and allows a direct determination of the lowest
order corrections to their  anomalous dimensions and OPE coefficients.
The choice obviously falls on the operators in the ${\cal N}$=4 
current multiplet. They play a central r\^ole in 
the correspondence with type IIB superstring theory on 
$AdS_{5}\times S^{5}$, since they couple to the fields 
of the gauged supergravity multiplet. 
Differently from AdS-inspired computations, where the scalar singlets of the 
$SU(4)$ R-symmetry in 
the dilaton-axion sector are amenable to explicit 
computations~\cite{dfmmr}, 
we prefer to work, as in~\cite{bgkr,dhkmv} and in~\cite{dfs}, with the
lowest scalar components in the current supermultiplet, ${\cal 
Q}_{{\bf 20}}^{ij}$, that belong to the representation ${\bf 
20}$ of $SU(4)$.  In terms of the six real scalars $\varphi^i$
belonging to the representation  ${\bf{6}}$ of $SU(4)$, they are defined as  
\begin{equation}
{\cal Q}_{{\bf 20}}^{ij} = \tr( \varphi^i \varphi^j - {\delta^{ij} 
\over 6} 
\varphi_k \varphi^k) \; . 
\end{equation} 
The trace over the $SU(N)$ colour indices, $a,b = 
1,\ldots, N^{2}-1$, is defined as usual by 
\be
\tr(T^{a} T^{b}) = {1 \over 2 } \delta^{ab} \; ,
\label{tracedef}
\ee
where $T^{a}$ are the generators in the fundamental representation of 
the $SU(N)$  gauge group. In order to fix our normalisations we  
write down explicitly the two-point function
of the scalar fields, which reads  
\begin{equation} 
\langle \varphi^{ia}(x_1) \varphi^{jb}(x_2)\rangle =  
\frac{1}{(2 \pi)^2} \frac{\delta^{ij}\delta^{ab}}{x_{12}^2} \; , 
\label{twopoint} 
\end{equation}
where $x_{pq}=x_p-x_q$. 

In order to analyse the short-distance behaviour of 4-point functions,
it is convenient to start from the OPE of two 
${\cal Q}_{{\bf 20}}$. At tree level it takes the form
\begin{eqnarray}
     {\cal Q}_{\bf{20}}(x_1)
    {\cal Q}_{\bf{20}}(x_2)  &=& 
    \frac{N^2-1}{(2\pi)^{4}x_{12}^{4}} + \frac{1}{(2\pi)^{2}x_{12}^{2}}
    \left[ {\cal Q}_{\bf{20}}({x_1+x_2 \over 2}) + 
{\cal K}_{\bf{1}}({x_1+x_2 \over 2})
\right]
    + \nonumber \\
    && + \frac{{x_{12}}_{\mu}}{(2\pi)^{2}x_{12}^{2}}\left[
    {\cal J}^{\mu}_{\bf{15}}({x_1+x_2 \over 2}) +
    {\cal K}^{\mu}_{\bf{15}}({x_1+x_2 \over 2}) \right] + \ldots \; ,
    \rule{0pt}{17pt} \label{qope}
\end{eqnarray}
where track has been kept of primary fields only and 
all the $SU(4)$ indices are omitted. The dots represent
operators of (tree-level) scale dimension $\Delta^{^{(0)}}\geq 4$. Bold
subscripts denote the $SU(4)$ representations to which the various operators 
belong. 
Besides the identity operator, there appear protected as well as {\it a priori} 
unprotected operators. 
Among the protected single-trace operators one finds ${\cal 
Q}_{\bf{20}}$ itself with $\Delta =2$, and the R-symmetry current 
${\cal J}^{\mu}_{\bf{15}}$ with $\Delta = 3$. 
Among the unprotected single-trace operators one finds 
operators belonging to the long Konishi multiplet, which we denote by ${\cal
K}$. ${\cal K}$ starts with the  $SU(4)$
singlet scalar of naive dimension $\Delta^{^{(0)}} = 2$ 
\begin{equation} 
{\cal K}_{{\bf 1}} = \frac{1}{3} : \tr  (\varphi^{i} 
\varphi^{i}) : \; 
\label{k1}
\end{equation}
and includes also the classically conserved 
current, ${\cal K}^{\mu}_{\bf{15}}$ with $\Delta^{^{(0)}} = 3$. As discussed at 
the end of sect.~2, positivity constraints on anomalous dimensions imply the
bound  $\Delta \ge 2$ for the dimension of ${\cal K}_{\bf{1}}$.

In order to illustrate 
the origin and the r\^ole of short distance 
logarithmic singularities 
 in the simplest way, it is convenient to
consider the four-point Green function
\begin{equation}
	G^{(H)}(x_{1}, x_{2}, x_{3}, x_{4}) = 
	\langle {\cal C}^{11}(x_1)  {\cal C}^\dagger_{11}(x_2)  
	{\cal C}^{22}(x_3) {\cal C}^\dagger_{22}(x_4)\rangle \; 
	\label{h4point}
\end{equation}
where we have introduced the gauge-invariant composite operators
\begin{equation}
{\cal C}^{IJ} = \tr (\phi^{I}\phi^{J}) \quad \quad  
{\cal C}^\dagger_{IJ} = \tr (\phi^{\dagger}_{I}\phi^{\dagger}_{J}) 
\; .
\label{defchir}
\end{equation}
In the previous equation we have defined 
the complex elementary fields 
\begin{equation}
\phi^{I} = { 1\over\sqrt{2}}\left( \varphi^I + i \varphi^{I+3}\right)
\quad \quad
\phi^\dagger_{I} = { 1\over\sqrt{2}}\left( \varphi^I - i 
\varphi^{I+3}\right) 
\qquad I=1,2,3 \; ,
\label{phinonvar}
\end{equation}
that belong to the representations ${\bf 3}_{+1}$ and ${\bf 3}_{-1}$ in the 
decomposition of the  representation ${\bf 6}$ of $SU(4)$ with respect to 
$SU(3)\times U(1)$.

At tree-level $G^{(H)}$ contains only disconnected diagrams and 
has the expression 
\begin{equation}
	G^{(H)}_{_{(0)}}(x_{1}, x_{2}, x_{3}, x_{4})  = 
\frac{ (N^2 - 1)^2}{4 (2\pi)^{8}{x_{12}^4 x_{34}^4}} \; . 
	\label{tree4point}
\end{equation}
From the point of view of the OPE~(\ref{qope})
 the $x$-dependence of 
eq.~(\ref{tree4point}) may appear very surprising, because one would 
naively expect that intermediate operators  
of dimension 2, 3  etc. should contribute giving rise to additional 
subdominant singularities in $x_{pq}$.
The point is that the Green function~(\ref{h4point}) 
can be expressed as the product of the two OPE's
\begin{equation}
 {\cal C}^{11}(x) {\cal C}^\dagger_{11}(y)  = 
 \frac{N^2-1}{2(2\pi)^{4}(x-y)^4} + \frac{1}{(2\pi)^{2}(x-y)^2}
 \left({\cal K}_{\bf{1}} + {\cal Q}^{(Y)}_{\bf{20}} +
 {\cal Q}^{(X)}_{\bf{20}}\right) 
 \left(\frac{x+y}{2}\right) + \ldots \; 
 \label{qqbarope1} 
\end{equation}
and 
\begin{equation}
 {\cal C}^{22}(x) {\cal C}^\dagger_{22}(y)  = 
 \frac{N^2-1}{2(2\pi)^{4}(x-y)^4} + \frac{1}{(2\pi)^{2}(x-y)^2}
 \left({\cal K}_{\bf{1}} + {\cal Q}^{(Y)}_{\bf{20}} -
 {\cal Q}^{(X)}_{\bf{20}}\right) 
 \left(\frac{x+y}{2}\right) + \ldots \; ,
 \label{qqbarope2} 
\end{equation}
where
\begin{eqnarray}
   {\cal Q}^{(Y)}_{\bf{20}} &=&  
   \frac{1}{3}  \tr(\phi^\dagger_{1}\phi^{1} + 
   \phi^\dagger_{2}\phi^{2} - 2 \phi^\dagger_{3}\phi^{3}) 
   \label{decompq1} \\
   {\cal Q}^{(X)}_{\bf{20}} &=& 
     \tr(\phi^\dagger_{1}\phi^{1} - 
   \phi^\dagger_{2}\phi^{2}) \\
   {\cal K}_{1} &=& \frac{1}{3} : \tr  (\varphi^{i} 
\varphi^{i}) :  \ = \ 
     {2 \over 3} : \tr(\phi^\dagger_{I}\phi^{I})  : \; 
\label{decompq2}
\end{eqnarray}
and cancellations among
the contributions of different  operators take place. 
For example, the pole 
${1}/{x_{12}^{2}}$ in the limit $x_{12}\rightarrow 0$ ($s$-channel) 
is absent, since the contribution of 
the operator ${\cal Q}^{(X)}_{\bf{20}}$  
exactly cancels the contributions of ${\cal 
Q}^{(Y)}_{\bf{20}}$ and ${\cal K}_{\bf{1}}$ to the four-point 
function~(\ref{tree4point}). Note that this cancellation is not 
possible at higher
orders, since ${\cal K}_{\bf{1}}$ has an anomalous dimension while ${\cal
Q}_{\bf{20}}$ is protected.

Similar tree-level cancellations take place among the spin one currents. For
instance, the potential pole ${x_{12}^{\mu}}/{x_{12}^{2}}$ has vanishing 
coefficient because the components of the currents, 
${\cal K}^{\mu}_{\bf{15}}$ and ${\cal J}^{\mu}_{\bf{15}}$, that 
couple to ${\cal C}^{11}(x) {\cal C}^\dagger_{11}(y)$ are orthogonal 
to the ones that couple to ${\cal C}^{22}(x) {\cal C}^\dagger_{22}(y)$.
Notice that this is true separately for ${\cal K}^{\mu}_{\bf{15}}$ and 
${\cal J}^{\mu}_{\bf{15}}$, so that this cancellation will
survive radiative corrections 
despite the fact 
that they induce an anomalous dimension for 
${\cal K}^{\mu}_{\bf{15}}$ but not for ${\cal J}^{\mu}_{\bf{15}}$.
Anticipating the results of the next section, let us note that 
at one loop the only field of tree-level dimension 
$\Delta^{^{(0)}} \leq 10$ with non vanishing contribution to
the $s$-channel of the function~(\ref{h4point}) is the lowest  Konishi 
scalar ${\cal K}_{\bf{1}}$. This demonstrates that reading off the 
complete operator content from a given four-point function might be rather
subtle. 

Let us briefly comment on the OPE in the other two channels of 
$G^{(H)}_{_{(0)}}$ in~(\ref{tree4point}).
The limit $x_{13}\rightarrow 0$ in~(\ref{h4point}) exposes the 
$u$-channel (the $t$-channel is similar) and singles out the double-trace 
operator
\begin{equation} 
{\cal D}^{(11|22)}= {\cal C}^{11} {\cal C}^{22} \; .
\label{d1122}
\end{equation}
This operator is automatically normal ordered and belongs
to the reducible ${\bf 105}+{\bf 84}$ representation of $SU(4)$.
In fact one has
\begin{eqnarray}
{\cal D}^{(11|22)}_{\bf{105}} &=& {1\over 3} ({\cal C}^{11} {\cal C}^{22}  
+ 2 {\cal C}^{12} {\cal C}^{12}) \label{defd105} \\ 
 {\cal D}^{(11|22)}_{\bf{84}} &=& {1\over 2} ({\cal C}^{11} {\cal C}^{22}  
- {\cal C}^{12} {\cal C}^{12} ) \label{defd84} \; .
\end{eqnarray}

As indicated, the operator ${\cal D}^{(11|22)}_{\bf{105}}$ belongs to the 
${\bf 105}$ representation of $SU(4)$ whose highest weight is $[0,4,0]$. 
Since its naive dimension, $\Delta^{^{(0)}}$, 
is exactly 4, ${\cal D}^{(11|22)}_{\bf{105}}$ is 
the lowest scalar component of a short and thus protected supermultiplet. 
Without spoiling the $SU(2,2|4)$ superconformal invariance, the supermultiplet 
that starts with ${\cal D}_{\bf{105}}$ can be made orthogonal to the 
short supermultiplet associated to the quartic Casimir that starts
with the scalar composite operator 
\be
{\cal Q}_{\bf 105} = \tr (\phi^{4}) \; . 
\label{q4def}
\ee
In particular it is convenient to define the combination
\be
\widehat{\cal Q}_{\bf 105} \equiv  {\cal Q}_{\bf 105} - 
{\langle {\cal D}^{\dagger}_{\bf{105}} 
{\cal Q}_{\bf 105} \rangle 
\over \langle {\cal D}^{\dagger}_{\bf{105}} 
{\cal D}_{\bf 105} \rangle } 
{\cal D}_{\bf{105}} =  \tr (\phi^{4}) - {2 N^{2}- 3 \over 
N(N^{2}+1) } [\tr (\phi^{2})]^{2} \; ,
\label{qhat}
\ee
that vanishes for $N=2$ \ie for an $SU(2)$ gauge group. 
Notice that with this definition $\widehat {\cal Q}_{\bf 105}$ has 
zero three-point function with two ${\cal Q}_{{\bf 20}}$, for any $N$. 
As a consequence the operator $\widehat {\cal Q}_{\bf 105}$ 
cannot contribute to the short distance limits that we 
study. The only operator in the representation ${\bf 105}$ that is relevant 
for our analysis is thus ${\cal D}_{\bf 105}$. In the next sections we will 
argue that ${\cal D}_{\bf 105}$ does not receive quantum corrections to its 
naive scale dimension and OPE coefficients.
The above observation about the decoupling of 
$\widehat {\cal Q}_{\bf 105}$ is in line with its AdS interpretation 
as a Kaluza-Klein excitation ($\ell=4$) of ${\cal Q}_{\bf 20}$ and clarifies 
a point raised in~\cite{dfs,liutse}.

The operator ${\cal D}^{(11|22)}_{\bf{84}}$ belongs to a long 
supermultiplet instead since the highest 
weight of the ${\bf 84}$ has Dynkin labels $[2,0,2]$, while the naive 
dimension of ${\cal D}^{(11|22)}$ is $\Delta^{^{(0)}} = 4$. 
The operator ${\cal D}^{(11|22)}_{\bf{84}}$ mixes with
\begin{equation} 
{\cal K}^{(11|22)}_{\bf{84}} = g^2 \tr ([\phi^{1},\phi^{2}] 
[\phi^{1},\phi^{2}]) \; ,
\label{defk84}
\end{equation}
that lies in the ${\bf 84}$ of $SU(4)$ and belongs 
to the Konishi supermultiplet~\footnote{Notice 
that the presence of the factor $g^2$ in the 
normalisation of ${\cal K}^{(11|22)}_{\bf{84}}$ is dictated by the 
form of the supersymmetry transformations acting on the lowest 
component ${\cal K}_{{\bf 1}}$ in eq.(\ref{k1}) and implies a 
shortening of the Konishi supermultiplet in the free field-theory 
limit~\cite{af}.}. The mixing is maximal for $SU(2)$ 
in which case ${\cal K}_{\bf{84}}$ and ${\cal D}_{\bf{84}}$ are 
proportional to one another. In general it is convenient to 
define an operator $\widehat{\cal D}_{\bf 84}$ that has vanishing 
two-point function at tree-level with ${\cal K}_{\bf{84}}$ much in the same 
way as in~(\ref{qhat}). At higher order one should appropriately 
adjust the form of the linear combinations that diagonalise the two-point 
functions in order to construct operators with well-defined anomalous 
dimensions.

\section{One-loop perturbative calculations}

In perturbative calculations it is necessary to work with an off-shell 
formulation, so that an ${\cal N}$=4 superfield approach, which admits
only an on-shell description, cannot be employed. The use of ${\cal N}$=1 
superspace techniques greatly simplifies the necessary algebra 
compared to component calculations. 
Alternatively one could resort to the manifestly ${\cal N}$=2 
supersymmetric harmonic superspace approach, as in~\cite{ehssw}, in 
which the ${\cal N}$=4 multiplet is realised putting together a 
${\cal N}$=2 vector multiplet and a hypermultiplet.

We prefer to work in the more familiar ${\cal N}$=1 formalism.
In this approach, the ${\cal N}$=4 vector multiplet 
decomposes into three ${\cal N}$=1 chiral multiplets ${\Phi^{I}}$ and 
one ${\cal N}$=1 vector multiplet, all in the adjoint representation of 
the $SU(N)$ gauge group. The lowest scalar components of the chiral 
multiplets are the ${\phi^{I}}$ introduced above.

Let us start by computing the superspace extension of $G^{(H)}(x_{1}, 
x_{2}, x_{3}, x_{4})$
that we denote by $\Gamma^{(H)}(z_{1}, z_{2}, z_{3}, z_{4})$, 
where
$z_{p}=(x_{p}, \theta_{p}, \bar\theta_{p})$ are the coordinates 
of ${\cal N}$=1 superspace. As in the component field computations
of~(\ref{h4point}), the only tree level contribution to
\begin{equation}
	\Gamma^{(H)}(z_{1}, z_{2}, z_{3}, z_{4}) = 
	\langle 
	\tr\Big[\big({\Phi^{1}}\big)^{2}\Big] (z_1)  
	\tr\Big[\big({\Phi_{1}^{\dagger}}\big)^{2}\Big] (z_2)
	\tr\Big[\big({\Phi^{2}}\big)^{2}\Big] (z_3)
	\tr\Big[\big({\Phi_{2}^{\dagger}}\big)^{2}\Big] (z_4)
	\rangle 
	\label{h4super}
\end{equation}
comes from a disconnected super-diagram that grows as $N^{4}$. Because of
colour contractions, the order $g^2$ correction to the disconnected diagram 
is trivially zero. In the Fermi-Feynman gauge, no corrections to the 
superfield propagators should be included~\cite{kov} and there is a  unique
non-vanishing first order correction represented by a single  connected
diagram, in which the internal propagator corresponds to the exchange of a
chiral  superfield with ``flavour'' $I=3$. Moreover, because of the
non-renormalisation theorems  for the two-point functions discussed
in~\cite{dfs}, there  are no non-vanishing disconnected super-diagrams
contributing  at first or higher order. 

Since we are dealing with Green functions of composite operators the 
result is necessarily depends on the $\theta$-variables associated to 
each of the four external points. This is not in 
agreement with the results of calculations presented 
in~\cite{gonzrparkschalm}. 
However, if we restrict our attention to the lowest components of the
chiral superfields, the expression in~\cite{gonzrparkschalm}, though 
rather implicit, turns out to be correct. The explicit result is 
\be
G_{_{(1)}}^{(H)}(x_{1}, x_{2}, x_{3}, x_{4}) = - { g^2 N (N^2-1) 
\over
2 (2 \pi)^{12}  x_{12}^2 x_{34}^2} \ \int d^4 x_0
{1 \over x^2_{10} x^2_{20} x^2_{30} x^2_{40}}  \; .  
 \label{gq4i} 
\ee
Introducing Feynman parameters it can be expressed as 
\be
G_{_{(1)}}^{(H)}(x_{1}, x_{2}, x_{3}, x_{4}) = - { g^2 N (N^2-1) \pi^2 
\over
2 (2 \pi)^{12}  x_{12}^2 x_{34}^2 x_{13}^2 x_{24}^2} \  B(r,s) \; ,  
 \label{gq4} 
\ee
where $B(r,s)$ is a box-type integral given by
\be 
B(r,s) = \int {d \beta_0 d \beta_1 d \beta_2
 \ \delta(1-\beta_0 - \beta_1 -\beta_2) \ 
 \over 
\beta_1 \beta_2 + r \beta_0 \beta_1  + s \beta_0 \beta_2 }
\; .
\label{Brs} 
\ee
As indicated, $B(r,s)$ depends only on the two
independent conformally invariant cross ratios 
\be
	r = {x_{12}^{2}x_{34}^{2} \over x_{13}^{2}x_{24}^{2}}
	\;  , \quad
	s = {x_{14}^{2}x_{23}^{2} \over x_{13}^{2}x_{24}^{2}}
	\; .
\label{crossrat}
\ee
The result of the integration in~(\ref{Brs}) can be expressed as a 
combination 
of logarithms and dilogarithms as follows
\ba
B(r,s) & = & 
{1 \over \sqrt{p}}
\left \{ \ln (r)\ln (s)  - 
\left [\ln \left({r+s-1 -\sqrt {p} \over 2}\right)\right ]^{2} + 
\right. \nonumber \\ 
&& \left. -2 {\rm Li}_2 \left({2 \over 1+r-s+\sqrt {p}}\right ) -
2 {\rm Li}_2 \left({2 \over 1-r+s+\sqrt {p}}\right )\right \} 
\; ,  \label{Brsf}
\ea
where ${\rm Li}_2 (z) = \sum_{n=1}^{\infty} {z^n\over n^2}$ 
and~\footnote{Notice the change of notation with respect to~\cite{bgkr}:
$\Delta$ there is $p$ here.} 
\be 
p = 1 + r^{2} + s^{2} - 2r - 2s - 2rs \; .  
\label{pdef}
\ee
Unlike correlation functions of elementary SYM fields, that are 
infra-red 
problematic and gauge-dependent (see \eg~\cite{kov,phd} for a recent 
analysis),
the correlator~(\ref{gq4}) is well defined at non-coincident points
and it has all the expected symmetry properties.

Notice that the representation of $B(r,s)$ given by~(\ref{Brsf})
is valid only  in the `physical 
domain', \ie in the region resulting from physically allowed choices 
of $x_{pq}^{2}$. In the Euclidian regime it is  defined by the condition 
$p \leq 0 $. Although individual terms in~(\ref{Brsf}) 
are complex, the complete function is real and positive in the 
`physical domain'. Moreover the singularities corresponding to the 
three possible 
channels in the four-point function $G^{(H)}$  
are located at $(r=0,s=1)$, $(r=1,s=0)$ and  $(r=\infty,s=\infty)$.

Let us consider  
the limit $x_{12}\rightarrow 0$. The four-point 
function behaves as
\begin{equation}
 x_{12}\rightarrow 0 : \quad G_{_{(1)}}^{(H)} (x_{1}, x_{2}, x_{3}, x_{4}) 
\rightarrow 
 { g^2 N (N^2-1) \pi^2 \over
  (2 \pi)^{12}  x_{12}^2 x_{34}^2 x_{13}^2 x_{24}^2 } 
 \left( {1 \over 2 } \log{x_{12}^2 x_{34}^2 \over x_{13}^2 x_{24}^2} 
-1 
 \right) \; .
 \label{oneloopuchlog1}
\end{equation}
According to eq.~(\ref{opex}), we can extract the one-loop 
contribution to the anomalous dimension of the Konishi multiplet
as the coefficient of the logarithmic term, after dividing out 
the tree-level contribution 
\begin{equation}
    \langle {\cal K}_{{\bf 1}}(x) {\cal K}_{{\bf 1}}(y) 
\rangle_{_{(0)}}  =
  {N^{2}-1 \over 3 (2\pi)^{4} (x-y)^{4}} \; 
 \label{twokonishi}
\end{equation}
and taking into account the tree-level result
\begin{equation}
  \langle {\cal C}^{11}(x_{1}) {\cal C}^\dagger_{11}(x_{3}) 
  {\cal K}_{{\bf 
  1}}(x_{2}) \rangle_{_{(0)}}  =
  {N^{2}-1 \over 3 (2\pi)^{6} (x_{13})^{2}(x_{12})^{2}(x_{23})^{2}} 
\; .
 \label{threekonishi}
\end{equation}
We get in this way
\begin{equation}
\gamma_{\cal K}^{\rm{1-loop}}(g) = {3 \over 16 \pi^2} {g^2 N} \; ,
\label{oneloopkoni}
\end{equation}
in agreement with previous results~\cite{ans}.
Indeed the only other gauge-invariant operators of  dimension two
that can be exchanged are the protected operators ${\cal Q}_{\bf 20}$, 
which cannot receive quantum corrections to their scale dimensions.

The subleading logarithmic singularity in this channel involves a rather 
complicated mixture of double-trace operators and superconformal 
descendants of the Konishi multiplet that will be discussed in detail 
in~\cite{futw}.

Let us briefly comment on the other channels. The limit $x_{13}\rightarrow 0$ 
in~(\ref{gq4}) exposes the 
$u$-channel (the $t$-channel is similar) and singles out the double-trace 
operator ${\cal D}^{(11|22)}$ defined in~(\ref{d1122}).
As already observed, the ${\cal D}_{\bf 105}$ component is protected 
from receiving corrections to OPE coefficients and anomalous dimension.
This can be seen more clearly by considering the short distance 
behaviour of the one-loop contribution to the four-point function 
considered in~\cite{bgkr}.
The presence of logarithms in this channel imply that one or both operators 
of naive dimension $\Delta^{^{(0)}}=4$ in the ${\bf 84}$ in eqs.~(\ref{defd84}), 
(\ref{defk84}) have anomalous 
dimension. Although ${\cal K}_{{\bf 84}}$ alone could 
account for the logarithm in this channel, leaving  
$\widehat{{\cal D}}_{{\bf 84}}$ with 
vanishing anomalous dimension at one-loop level, in order to settle 
this issue one has to consider another independent Green function~\cite{futw}.  
For instance one can take the 
$\theta_{1}^{2}{\overline \theta}_{2}^{2}\theta_{3}^{2}
{\overline \theta}_{4}^{2}$-component of the Green function~(\ref{h4super})
\be
\langle {\cal E}^{11}(x_{1}){\cal E}^{\dagger}_{11}(x_{2})
{\cal E}^{22}(x_{3}){\cal E}^{\dagger}_{22}(x_{4}) \rangle \; , 
\label{e4}
\ee
where, denoting by $\psi^{I}$ and $F^{I}$ the fermionic and auxiliary 
components of the $\Phi^{I}$ superfield, we have introduced the 
definition 
\be
{\cal E}^{{IJ}} = \tr (\psi^{I} \psi^{J} + 
\phi^{I}F^{J}+\phi^{J}F^{I}) \; .
\label{edef}
\ee
${\cal E}$ has protected dimension $\Delta=3$ and belongs to the 
representation 
${\bf 6}_{+1}$ in the decomposition of the ${\bf 10}$ of $SU(4)$ with 
respect to $SU(3)\times U(1)$.

We would like to conclude this section with a general observation
about the large $N$-behaviour of the anomalous dimensions of
unprotected double-trace  operators ${\cal D}$,
which schematically we write  as 
\be
{\cal D} = \; : \tr( \varphi \varphi ) \; 
\tr( \varphi \varphi )  : \; .
\ee
Their one-loop anomalous dimensions are expected to be of the form  
\be
\gamma_{{\cal D}}^{\rm{1-loop}}(g) = {c_N^{(1)}\over 
4\pi^{2}}{ g^2 N \over N^2} \; ,
\label{dtoneloop}
\ee
where $c_N^{(1)}$ is a (possibly vanishing) constant of O(1)
for large $N$. This follows by looking at the large $N$-behaviour of
their tree-level two-point function
\be
\langle {\cal D}^\dagger(x){\cal D}(y) \rangle_{_{(0)}} = {a^{(0)}_N 
\over (2\pi)^8} {N^4 \over (x-y)^8} \; , 
\label{anothertwo}
\ee
where again $a^{(0)}_N$ is a positive constant of O(1) in the large
$N$ limit, which depends on the specific choice of the operator. 
We expect the same pattern to persist at higher 
$L$-loops with
\be
\gamma_{{\cal D}}^{L-{\rm loop}}(g) = {c_N^{(L)}\over 
(4\pi^{2})^{L}}{(g^2 N)^L 
\over N^2}
\label{dtelloop}
\ee
implying a vanishing of the perturbative value of 
$\gamma_{{\cal D}}$ in the large $N$ limit at all orders.

Similar arguments apply to multi-trace operators, ${\cal M}$, for 
which one finds
\be
\langle {\cal M}^\dagger(x){\cal M}(y) \rangle_{_{(0)}} = {b^{(0)}_N 
\over (2\pi)^{4q}}
{N^{2q} \over (x-y)^{4q}} \; , 
\label{multitwo}
\ee
where $q$ denotes the number of colour traces in the definition of 
${\cal M}$ and $b^{(0)}_N$ is a positive constant of O(1) for large 
$N$.

\section{Non-perturbative instanton calculations}

Relying on the remarkable properties of instanton calculus, the 
one-instanton contribution to a four-point function 
in ${\cal N}$=4 SYM theory with $SU(2)$ gauge group has been computed (to
lowest order in the  coupling constant) in ref.~\cite{bgkr}. The extension of
that result  to the case of $SU(N)$ and to any instanton number in the large 
$N$ limit has been given in~\cite{dhkmv}.

Following the steps detailed in~\cite{bgkr}, we give here  
a closed-form expression for the one-instanton 
contribution to  $G^{(H)}$ (eq.~(\ref{h4point})).

We recall that in order to saturate the sixteen 
`unlifted'~\cite{dhkmv}
fermionic zero-modes present in the one-instanton background, the 
fields 
$\phi^{I}$ and ${\phi}^\dagger_{I}$ have to be `replaced' 
with the expressions 
\begin{eqnarray} 
&& \phi^{I}(x) \rightarrow \phi^{I}_{_{(0)}}(x) =  \frac{1}{2\sqrt{2}}
\zeta^{0} \sigma^{\mu\nu} \zeta^{I}   
{F}_{\mu\nu}  \nonumber \\ 
&& {\phi}^{\dagger}_{I}(x) \rightarrow 
{\phi}^\dagger_{I_{(0)}}(x) =  
\frac{1}{4\sqrt{2}} \varepsilon_{IJK} \zeta^{J} 
\sigma^{\mu\nu} \zeta^{K} { F}_{\mu\nu} \; , 
\label{scalarsol} 
\end{eqnarray}  
where ${ F}_{\mu\nu}$ is the one-instanton field-strength and 
$\zeta^{A}_{\alpha}=\eta^{A}_{\alpha}+\widehat{x}_{\alpha{\dot\alpha}}
{\overline\epsilon}^{^{\dot\alpha}A}$, $A=I,0$, 
with 
$\widehat{x}_{\alpha{\dot\alpha}}=x_{{\mu}}(\sigma^{\mu})_{\alpha 
\dot\alpha}$, are fermionic collective coordinates. The index $I=1,2,3$ 
labels the fermionic zero modes 
in the ${\cal N}$=1 chiral multiplets, the index $0$ those associated 
to the ${\cal N}$=1 gaugino. The instanton-induced `scalar 
fields'~(\ref{scalarsol}) satisfy the scalar equation in the 
presence of the fermionic zero-modes. They correspond to the leading 
non-vanishing effective field configurations that result from the Wick 
contractions with the Yukawa terms obtained after expanding the 
exponential of the action until a number of fermion fields equal to 
the number of instanton induced fermionic zero modes are brought down.
The scalar field configurations~(\ref{scalarsol}) can also be 
obtained directly from supersymmetry transformations 
by acting twice on ${F}_{\mu\nu}$ with the 
supersymmetry and superconformal transformations 
associated to $\zeta$'s, not preserved in the instanton background. 

After some Fierz transformations on the fermionic  
collective coordinates the fermionic integrations can be performed in 
a standard way~\cite{akmrv} and yield
\begin{eqnarray}  
&& G^{(H)}(x_{1}, x_{2}, x_{3}, x_{4}) = \nonumber\\
&& =\frac{3^{4}}{4\pi^{10}}\, g^8
\,e^{-{8\pi^2 \over g^2}} (e^{i\theta} + e^{-i\theta}) \,    
x_{13}^2 x_{24}^2 x_{14}^2 x_{23}^2 \, \int {d\rho_0 d^4 x_{0} \over 
\rho_0^5} \, \prod_{p=1}^4 {\left[ {\rho_0 
\over {\rho_0^2 + (x_p - x_{0})^2}}  \right]}^4 \; ,
\label{collectint} 
\end{eqnarray}
where we have added up instanton and anti-instanton 
contributions~\footnote{It should be noted that for this class of 
four-point functions instanton and anti-instanton contributions are simply 
complex conjugate to one another. On the contrary, at lowest order 
in $g$, 
correlation functions of protected operators 
that violate the external automorphism 
$U(1)_{B}$ of the ${\cal N}$=4 superconformal algebra, whose 
AdS counterpart is the anomalous $U(1)_{B}$ 
symmetry of the type IIB supergravity, only receive contribution from 
instantonic configurations  with topological charge of a given 
sign~\cite{bgkr,intri}.}.
Up to an overall conformal invariant factor  ${x_{14}^2 x_{23}^2 
\over x_{13}^2 x_{24}^2}$, the result~(\ref{collectint}) coincides 
with the one found in~\cite{bgkr}. 

Apart from numerical factors, the same results is obtained  for any 
$SU(N)$ gauge group~\cite{dhkmv}. This follows from the fact that 
the extra `non-geometric' $8N-16$ fermionic zero-modes, present 
for $N > 2$, turn out to be effectively lifted, as they appear 
in quadrilinear terms in the full SYM action. The lifting can be most 
easily understood as the result of the functional integration over 
the scalar field fluctuations around the classical configuration 
($\phi=0$), noticing that, in the semiclassical approximation in 
which we are working, fermionic fields are to be effectively 
replaced by their zero mode expression. Bosonising the lifted 
fermionic zero-modes {\it \'a la} Hubbard-Stratonovich 
and performing the extra bosonic integrations associated to the 
new bosonic zero modes produces the extra factor
$\Gamma(N-{1\over2}) / \Gamma(N-{1})$ in front 
of~(\ref{collectint}) that precisely matches the AdS-inspired 
expectation, $\sqrt{N}$, in the large $N$ limit~\cite{dhkmv}.

Higher instanton numbers may be exactly taken into account in the 
large $N$ limit, because a saddle-point approximation is possible.
Contrary to naive expectations, the instanton gas is far from being 
dilute in the sense that at large $N$ the $K$ instantons tend to 
``attract'' each other and share the same position, $x_0$, and size, 
$\rho_0$. Still they lie in $K$ commuting $SU(2)$ subgroups of the 
$SU(N)$ gauge group. Properly taking into account the r\^ole of the 
lifted and bosonised fermionic zero-modes, one can show that the 
situation is as if the $K$ instanton moduli space were a single 
copy of $AdS_5\times S^5$~\cite{dhkmv}. 
The same situation prevails in type IIB D-instanton 
computations and all $K$-dependent factors on the SYM side
felicitously reproduce the type IIB D-instanton induced 
terms~\cite{gg,bg,bgkr}.

In summary, up to computable overall constants, the one-instanton contribution 
to the four-point functions under consideration is the same for any 
$SU(N)$ gauge group and, in the large $N$ limit, every instanton 
number contributes the same expression. In the large $N$ limit, the 
sum over $K$ gives rise to an overall non-holomorphic function of the 
complexified gauge coupling $\tau$, which coincides with the 
non-perturbative D-instanton contribution to the coefficient,
$f_4(\tau,\bar\tau)$,  appearing in front of the ${\cal R}^4$ terms in the
type IIB low-energy effective action~\cite{gg}. 

Going back to eq.~(\ref{collectint}), we notice that the $x_{0}$ integration 
resembles that of a standard 
Feynman diagram with momenta replaced by position differences and can 
be performed either by expressing it in terms of derivatives of the 
box-integral as
\begin{equation}  
\int {d\rho_0 d^4 x_{0} \over \rho_0^5} 
 \, \prod_{p=1}^4 {\left[ {\rho_0 \over 
{\rho_0^2 + (x_p - x_{0})^2}}  \right]}^4 = \frac{5\pi^{2}}{108} 
\prod_{p<q} {\partial \over \partial x_{pq}^2} 
\int \prod_{p} d\alpha_p  
{\delta \left( 1-\sum_q \alpha_q \right)  
\over (\sum_{p,q} \alpha_p \alpha_q x_{pq}^2 )^{2}} \; , 
\label{intbox} 
\end{equation}
or by introducing the alternative parameterisation
\be
\int {d\rho_0 d^4 x_{0} \over \rho_0^5} 
 \, \prod_{p=1}^4 {\left[ {\rho_0 \over 
{\rho_0^2 + (x_p - x_{0})^2}}  \right]}^4  =  {5 \pi^2 \over 3}
{1 \over x_{13}^8 x_{24}^8} I(r,s) \; ,
\ee
where $I$ is the integral
\be 
I(r,s) = \int d \beta_0 d \beta_1 d \beta_2 
{\beta_0^3 \beta_1^3 \beta_2^3 \ \delta(1-\beta_0 - \beta_1 -\beta_2) 
\over 
(\beta_1 \beta_2 + r \beta_0 \beta_1  + s \beta_0 \beta_2 )^4}
\; .
\label{i1}
\ee
After a rather lengthy calculation both approaches yield 
\ba
I(r,s) & = & 
{P_0(r,s) \over {p}^{6}} \ B(r,s) +
{ P_1(r,s) \over 3 {p}^{6}}\ \ln (r)  +
{ P_1(s,r)  \over 3 {p}^{6}}\ \ln (s) + 
{ P_2(r,s) \over 18 {p}^{5}} \; , 
\label{nonpertf}
\ea
where $B(r,s)$ is given by~(\ref{Brsf}), $p$ is defined 
in~(\ref{pdef}) 
and $P_i(r,s)$ are polynomials in the cross ratios $r$ and $s$ 
introduced in~(\ref{crossrat}). Their explicit expressions are
\ba
 P_0(r,s) & = &
1+3 s+3 r-36 {r}^{5}{s}^{4}-498 r{s}^{5}+3948 {r}^{3}{s}^
{3}-498 {r}^{5}s-1338 {r}^{3}{s}^{4}+62 {r}^{3}{s}^{6} +
\nonumber \\ &&
-36 {r}^{4}{s}^{5}
+1050 {r}^{4}{s}^{4}+1512 {r}^{2}{s}^{2}+3 {r}^{8}s-30 {r}^{
2}{s}^{7}-30 {r}^{7}{s}^{2}-1338 {r}^{4}{s}^{3} +
\nonumber \\ &&
+1512 {r}^{5}{s}^{2}
+1512 {r}^{2}{s}^{5}+1050 {r}^{4}s-498 {r}^{5}{s}^{3}-498 {r}^{3}{
s}^{5}+62 {r}^{6}{s}^{3}+3 r{s}^{8} +
\nonumber \\ &&
-144 {r}^{6}s-144 {s}^{6}r-144
 {s}^{2}r+1050 r{s}^{4}-498 {r}^{3}s+114 r{s}^{7}-1338 {r}^{2}{s}
^{3} +
\nonumber \\ &&
-144 {r}^{6}{s}^{2}+114 {r}^{7}s-144 {r}^{2}{s}^{6}-498 r{s}^{
3}-1338 {r}^{4}{s}^{2}-144 {r}^{2}s-1338 {r}^{3}{s}^{2} +
\nonumber \\ &&
+114 rs- 1338 {r}^{2}{s}^{4}-30 {s}^{2}+62 {r}^{3}-36 {r}^{5}-
36 {r}^{4}+3{s}^{8}-30 {s}^{7}+62 {s}^{3} +
\nonumber \\ &&
-30 {r}^{7}+3 {r}^{8}+{s}^{9}+{r}^{9}+62 {r}^{6}+62 {s}^{6}-
36 {s}^{5}-36 {s}^{4}-30 {r}^{2} 
\nonumber
\ea
\ba
P_1(r,s) & = &
11-28 s+62 r-2142 r{s}^{5}+5132 {r}^{3}{s}^{3}+1170
 {r}^{5}s-4894 {r}^{3}{s}^{4}+215 {r}^{4}{s}^{4} +
\nonumber \\ &&
+4491 {r}^{2}{s}^{
2}+3404 {r}^{4}{s}^{3}+1170 {r}^{5}{s}^{2}+972 {r}^{2}{s}^{5}+3404
 {r}^{4}s-646 {r}^{5}{s}^{3} +
\nonumber \\ &&
+362 {r}^{3}{s}^{5}-1180 {r}^{6}s+590{s}^{6}r-2142 {s}^{2}r+
 1490 r{s}^{4}-4894 {r}^{3}s+62 r{s}^{7} +
\nonumber \\ &&
-10264 {r}^{2}{s}^{3}+383 {r}^{6}{s}^{2}-34 {r}^{7}s-331 {r}^{2}{s}
^{6}+1490 r{s}^{3}-8982 {r}^{4}{s}^{2}+972 {r}^{2}s +
\nonumber \\ &&
+5132 {r}^{3}{s}^{2}+590 rs+4491 {r}^{2}{s}^{4}-52 {s}^{2}+
362 {r}^{3}-646 {r}^{5}+215 {r}^{4}+11 {s}^{8}+
\nonumber \\ &&
-28 {s}^{7}+284 {s}^{3}-34 {r}^{7}-22 {r}^{8}+383 {r}^{6}-52 {s}^{6}+
284 {s}^{5}-430 {s}^{4}-331 {r}^{2} 
\nonumber
\ea
\ba
P_2(r,s) &=&-114 s-114 r-114 r{s}^{5}+2404 {r}^{3}{s}^{3}-
114 {r}^{5}s+15402 {r}^{2}{s}^{2}+4734 {r}^{4}s +
\nonumber \\ &&
-4620 {s}^{2}r+4734 r{s}^{4}-4620 {r}^{3}s-4620 {r}^{2}{s}^{
3}-4620 r{s}^{3}-1281 {r}^{4}{s}^{2} +
\nonumber \\ &&
-4620 {r}^{2}s-4620 {r}^{3}{s}^{2}+4734 rs-1281 {r}^{2}{s}^{4}-
1281 {s}^{2}+2404 {r}^{3}-114 {r}^{5} +
\nonumber \\ &&
-1281 {r}^{4}+2404 {s}^{3}+193 {r}^{6}+193 {s}^{6}-
114 {s}^{5}-1281 {s}^{4}-1281 {r}^{2}+193 \; .
\nonumber
\ea

Having obtained a closed expression for $G^{(H)}$, 
one can unambiguously study the behaviour when any two points are taken 
close to one another, $x_{pq}\to 0$.  An explicit calculation 
shows that $I(r,s)$ has singularities  only
at the points $(r=0,s=1)$, $(r=1,s=0)$ and $(r=\infty,s=\infty)$.
Taking into account the factor, 
$x_{13}^2 x_{24}^2 x_{14}^2 x_{23}^2$, present in the complete 
expression of $G^{(H)}$, one finds that the only case 
in which  there is actually a  singularity is when any
${\cal C}$ approaches its conjugate ${\cal C}^\dagger$,
which corresponds either to the limit $x_{12}\rightarrow 0$ or to
$x_{34}\rightarrow 0$. In terms of the conformal cross-ratios $s$ and 
$r$ introduced before, both limits, in fact, correspond to $r\to 0$, 
$s\to 1$, where $I(r,s)$ develops a logarithmic singularity. 
Notice that individual terms in $I(r,s)$ have pole-type singularities 
(of degree up to seven), but in the sum only a logarithmic singularity 
survives. 

In the limit $x_{12}\rightarrow 0$, the one-instanton 
plus one anti-instanton contribution to the four-point function 
behaves as
\begin{eqnarray}
 x_{12}\rightarrow 0: \quad && G^{(H)}(x_{1}, x_{2}, x_{3}, 
x_{4}) \rightarrow 
 \nonumber\\
&& \rightarrow \frac{15}{64\pi^{8}}\, g^8
\,e^{-{8\pi^2 \over g^2}} (e^{i\theta} + e^{-i\theta}) \,   
{ 1 \over x_{13}^4 x_{14}^4} \,
\left( - {2706\over 1225} - {36\over 35} 
\log{x_{12}^2 x_{34}^2\over x_{13}^2 x_{24}^2} \right) \; .
 \label{oneloopuchlog2}
\end{eqnarray}
The absence of a term of the type $\log(x_{12}^{2})/x_{12}^{2}$ 
implies that 
there is no one-instanton contribution to the  anomalous dimension of the
scalar ${\cal K}_{1}$. By supersymmetry arguments this result extends 
to the whole 
Konishi  multiplet. The absence of a term of the type $1/x_{12}^{2}$ confirms
the vanishing of any quantum correction to the two-point function of 
the protected 
gauge-invariant operators of dimension 2, ${\cal Q}_{\bf 20}$.
The leading logarithmic singularity present in~(\ref{oneloopuchlog2}) comes
from a mixture of double-trace operators of naive dimension
$\Delta^{^{(0)}}=4$. Extracting the non-perturbative contribution to the
anomalous dimension and OPE coefficients of all these operators is beyond
the  scope of this paper and will be discussed elsewhere~\cite{futw}.
Let us stress, however, that the non-vanishing coefficient of the 
logarithm in~(\ref{oneloopuchlog2}) implies that at least 
one double trace scalar operator of tree-level dimension 
$\Delta^{^{(0)}}=4$ has a non-vanishing one instanton correction 
to its anomalous dimension.

In the other two channels the vanishing of the factor 
$x_{13}^2 x_{24}^2 x_{14}^2 x_{23}^2$ 
leads to a vanishing result. This immediately implies that
only operators of dimension higher than four can contribute 
in these channels. 
Notice that performing a similar OPE analysis of the four-point function 
computed in~\cite{bgkr} leads one to conclude that ${\cal D}_{\bf 105}$ 
cannot be exchanged in the $u$-channel. The remaining operators of 
dimension four that can be exchanged in this channel
belong to the representation {\bf 84} of $SU(4)$
and are the component of the Konishi multiplet ${\cal K}_{\bf 84}$ in 
eq.~(\ref{defk84})
and the double-trace operator $\widehat{\cal D}_{\bf 84}$ introduced 
at the end of section 3.
Since, as we have demonstrated, the one-instanton correction to the anomalous 
dimension of ${\cal K}_{\bf 84}$ is zero, one can immediately 
conclude that the one-instanton correction to the anomalous dimension 
of $\widehat{\cal D}_{\bf 84}$ is zero as well.

Let us also note that in general the one (anti-)instanton contributions to
the  anomalous dimensions of double-trace operators 
are expected to be of the form
\begin{equation} 
\gamma_{\cal D}^{{\rm 1-inst}} =  {C^{K=\pm 1}_N\over N^4} 
{\Gamma(N-{1\over2})
\over  \Gamma(N-{1})} e^{-{8\pi^2 \over g^2}}(e^{i\theta} + e^{-i\theta}) \; ,
\end{equation} 
where $C^{K=\pm 1}_N$ is a constant of O(1) for large $N$. Once again 
the scaling of the denominator with $N^4$ is determined by the 
tree-level normalisation of the double-trace operators.

The observation that neither operators in the supercurrent
multiplet nor operators in the Konishi multiplet are exchanged in any 
of the intermediate channels at the one (anti-) instanton level, 
implies that the known non-renormalisation theorems for three-point functions
of protected operators~\cite{dfs} can be generalised at
non-perturbative level.
This result should not be considered in contradiction with the expected 
S-duality of the ${\cal N}$=4 SYM theory, as S-duality is not a pointwise
symmetry relating individual operators. For instance, string states, such as
those corresponding to the Konishi multiplet, get transformed into D-string
states or to other dyonic states, so that a significant reshuffling
among operators should take place under S-duality. Slightly at variant with 
respect to~\cite{intri}, we only expect that the whole spectrum of 
(anomalous) dimensions and the whole set of OPE coefficients should be 
invariant under S-duality. 

\section{Conclusions and comments on AdS}

The outcome of our analysis is that 
the short distance logarithmic behaviour of the Green functions in
${\cal N}$=4 SYM theory in the (super)conformal phase is
a manifestation of the presence of non vanishing anomalous dimensions
of unprotected operators.
A highly non-trivial, but technically very involved, check of the 
validity  of our interpretation, 
would consist in showing that the next  order 
corrections give rise to square logarithms with
coefficients that are precisely given by the squares of the first 
order coefficients as demanded for the exponentiation of the 
logarithmic terms. 

The presence of logarithms might be thought to spoil the perturbative 
finiteness properties of ${\cal N}$=4 SYM  theory that have been known for 
a long time. Although, as we have argued in this paper, this is not 
true, the analysis presented in~\cite{kov} has pointed out a number 
of problems in the perturbative computation of Green functions of 
elementary (super) fields. The calculations performed in~\cite{kov} 
have clarified many subtleties related to the gauge-fixing procedure 
of the theory, both in the description in terms of component fields 
and using the ${\cal N}$=1 superfield formalism. The appearance of 
UV divergences in off-shell propagators, when the  Wess--Zumino
(WZ) gauge is exploited, has been discussed and it 
has been shown that relaxing the WZ gauge choice makes UV divergences 
disappear. 

The analysis of~\cite{kov,phd} has also shown that in Green 
functions of elementary (super)fields there are IR divergences  
even for non-exceptional values of the momenta. These are in fact 
gauge artifacts and are due to the $1/k^{4}$ behaviour of the propagator 
of the lowest component of the vector superfield. 
This problematic term in the superfield propagator disappears in the 
(supersymmetric generalisation of the) Fermi--Feynman gauge and in 
the case the WZ gauge is employed. It would be interesting 
to obtain an explicit check of the cancellation of IR
divergences in gauge invariant Green functions for an arbitrary 
gauge-choice, even without restricting the theory to the WZ gauge.

Finally let us discuss the bearing of our results on the validity
and the meaning of the Maldacena conjecture. We have shown  by scaling
arguments that all the perturbative and non-perturbative contributions to
the anomalous dimensions  of multi-trace operators actually vanish in the
large $N$ limit. The situation for the Konishi multiplet is different, 
because its anomalous dimension does not seem to vanish in the large $N$ 
limit. Actually on the basis of the AdS/SCFT correspondence it is 
suggested that it grows as large as  $N^{1/4}$, eventually decoupling from 
the operator algebra. 

Short-distance logarithmic singularities have been shown to appear in 
genuine AdS supergravity calculations~\cite{dfmmr,sanjay}. These 
logarithms are expected to be related to the exchange of multi-particle
bound states in the bulk which are in correspondence with multi-trace 
operators on the boundary. In view of the vanishing of the anomalous 
dimensions of multi-trace operators that we find at large $N$, it is 
not obvious to us what is the relation between the logarithms seen in 
perturbative and non-perturbative ${\cal N}$=4 SYM calculations and 
the logarithms identified in~\cite{dfmmr,sanjay} despite their 
suggestive similarity~\cite{agmoo}. 

\vspace*{1cm}

{\large{\bf Acknowledgements}}

\vspace*{0.3cm}
\noindent
We would like to thank  L.~Andrianopoli, D.~Anselmi, M.~Bochicchio, 
S.~Ferrara, P.~Fr\'e, M.B.~Green, R.~Musto, F.~Nicodemi, R.~Pettorino, 
A.M.~Polyakov, D.~Polyakov, A.~Sagnotti, E.~Sokatchev, 
M.~Testa, I.T.~Todorov and 
K.~Yoshida for useful discussions. Ya.S.S. would like to thank 
the Physics Department and I.N.F.N. Section at Universit\`a di Roma 
``Tor Vergata'' for hospitality and financial support. The 
results of this investigation have been preliminarily presented by  
M.B. at the Meeting ``Advances in Theoretical Physics'' held in Vietri,
Italy, April 1999 and by S.K. at the ``Triangular 
Meeting on Quantum Field Theory in Particle Physics'' held in Utrecht, 
Holland, May 1999.

\end{document}